\documentclass[a4paper,twoside]{article}
\newcommand{\affil}[1]{$^{\rm #1}$}
\usepackage[authoryear]{natbib}
\bibpunct{(}{)}{;}{a}{}{,}
\usepackage{graphicx,psfig}
\date{} 
\newcommand{\kms}{\mbox{km\,s$^{-1}$}}
\newcommand\fs{\mbox{$.\!\!^{\mathrm s}$}}%
\newcommand\farcs{\mbox{$.\!\!^{\prime\prime}$}}%
\newcommand\Msun{\mbox{M$_{\odot}$}}%
\def\brg{\mbox{Br$\gamma$}}
\def\pab{\mbox{Pa$\beta$}}
\def\h2{\hbox{H$_{2}$ 1-0 S(1)}}
\newcommand\co{\mbox{CO(2,0)}}%
\newcommand\fe{\mbox{[Fe\,{\sc ii}]}}%
\newcommand\mic{\mbox{$\mu$m}}%

\title{\large\bf\flushleft The Nuclear Ring in the Barred Spiral Galaxy
IC~4933}
\author{\parbox{\textwidth}{\flushleft
\vspace{-0.5cm}
{\it Stuart D. Ryder\affil{A,D}, Samuel M. Illingworth\affil{B},
Robert~G.~Sharp\affil{A} and Catherine~L.~Farage\affil{C}}\\
\vspace{0.4cm}
{\small \affil{A}\,Anglo-Australian Observatory, P.O. Box 296, Epping,
NSW~1710, Australia}\\
{\small \affil{B}\,Department of Physics \& Astronomy, University of Leicester,
Leicester, LE1 7RH, UK}\\
{\small \affil{C}\,Research School of Astronomy and Astrophysics, Australian
National University, ACT 2611, Australia}\\
{\small \affil{D}\,Email: sdr@aao.gov.au}}}
\begin{document}
\twocolumn[
\begin{changemargin}{.8cm}{.5cm}
\begin{minipage}{.9\textwidth}
\vspace{-1cm}
\maketitle
%
%
\small{\bf Abstract:}

We present infrared imaging from IRIS2 on the Anglo-Australian Telescope
that shows the barred spiral galaxy IC~4933 has not just an inner
ring encircling the bar, but also a star-forming nuclear ring 1.5~kpc
in diameter. Imaging in the $u^{\prime}$ band with GMOS on Gemini
South confirms that this ring is not purely an artifact due to dust.
Optical and near-infrared colours alone however cannot break the
degeneracy between age, extinction, and burst duration that would
allow the star formation history of the ring to be unraveled.
Integral field spectroscopy with the GNIRS spectrograph on Gemini
South shows the equivalent width of the \pab\ line to peak in the
north and south quadrants of the ring, indicative of a bipolar
azimuthal age gradient around the ring. The youngest star-forming
regions do not appear to correspond to where we expect to find the
contact points between the offset dust lanes and the nuclear ring
unless the nuclear ring is oval in shape, causing the contact points
to lead the bar by more than $90^{\circ}$.

\medskip{\bf Keywords: stars: formation --- galaxies: evolution ---
galaxies: individual (IC~4933) --- galaxies: nuclei --- galaxies: starburst}

\medskip
\medskip
\end{minipage}
\end{changemargin}
]
\small

\section{Introduction}

High spatial resolution multi-wavelength imaging has revealed a host
of intriguing features close in to the nuclear regions of barred
spiral galaxies, including starburst rings, dust spirals, and nested
bars (e.g., \citealt{rk99}; \citealt{martini03}; \citealt{laine02}).
Early numerical simulations of dynamical evolution in barred galaxies (e.g.,
\citealt{sch81,ath82}) appeared to link circumnuclear starburst rings
with the inner Lindblad resonance (ILR), the radius at which the bar
pattern speed $\Omega_{p}$ is equal to $\Omega - \kappa / 2$, where
$\Omega$ is the mean angular frequency, and $\kappa$ the radial
epicyclic frequency. More recent hydrodynamic simulations by
\citet{rt03} have argued that star-forming nuclear rings actually result
whenever a certain fraction of gas clouds find themselves on elongated
orbits aligned perpendicular to, rather than parallel with, the major
axis of the bar, and that furthermore these rings shrink with time.

As atomic gas in the circumnuclear regions is generally scarce, the
bar (or some form of oval distortion) is thought to play a significant
role in `funneling' gas inwards via shocks to feed the competing needs
of a star-forming ring, and/or an Active Galactic Nucleus (AGN)
component. When gas is trapped near or between resonant orbits, and is
consumed in star formation, the AGN will potentially be starved and
forced to go dormant. One way to overcome this `logjam' is to create a
secondary, or `nested' bar, which can disrupt the resonances of the
main bar \citep{es04,hsa07}. Thus, we see that nuclear rings, nested bars,
and AGN are all fundamentally inter-related, and yet we have
surprisingly little data on their relative frequencies
\citep{jhk05}, or causal connections \citep{lkr06}.

The total sample of nuclear rings of sufficient angular size to be
resolved with ground-based seeing-limited spectrographs or aperture
synthesis interferometers is still quite limited
\citep{com10}. Galaxies exhibiting an inner ring encircling their bar
very often harbour a nuclear ring as well \citep{bc96}. The nearly
face-on barred spiral galaxy IC~4933 is listed in the {\em Catalog of
Southern Ringed Galaxies} (CSRG;
\citealt{csrg}) as having an inner ring of $47 \times 31$~arcsec across.
Table~\ref{t:ned} summarises some other properties of this galaxy. In
this paper, we report the discovery of an apparent nuclear ring in
IC~4933 from near-infrared (NIR) and optical imaging, and provide
spectroscopic confirmation of its nature. In Sections~\ref{s:iris2}
and \ref{s:gmos} we describe the image acquisition and reduction
procedures, and analyse the colour maps. We present follow-up NIR
Integral-Field Unit (IFU) spectroscopy in Section~\ref{s:gnirs} which
enables us to reconstruct the recent star formation
history. Section~\ref{s:disc} puts this in the context of an emerging
picture of sequential star formation being uncovered within nuclear
rings.

\begin{table}[h]
\begin{center}
\caption{Basic data for IC~4933}\label{t:ned}
\begin{tabular}{lcc}
\hline Parameter & Value & Source \\
\hline
Hubble Type & SB(rs)bc & 1 \\
RA (J2000)  & 20$^{\rm h}$03$^{\rm m}$29\fs04 & 2 \\
Dec (J2000) & $-54^{\circ}$58$^{\prime}$47\farcs8 & 2 \\
$V_{\rm hel}$ (km~s$^{-1}$) & 4915 & 3 \\
$D$ (Mpc)     & 66                   & 4 \\
Scale (pc arcsec$^{-1}$) & 313     & 4 \\
Inclination ($^{\circ}$) &  37     & 5 \\
Extinction $A_{B}$ (mag)  & 0.199 & 6 \\
\hline
\end{tabular}
\medskip\\
Sources: (1) \citealt{csrg}; (2) \citealt{2mass}; (3) \citealt{hicat};
(4) NED; (5) \citealt{lv89}; (6) \citealt{sch98}
\end{center}
\end{table}

\section{Observations and Results}

\subsection{Infrared Imaging}
\label{s:iris2}

Nuclear rings are seriously under-represented in the CSRG, not only
because the central regions of many galaxies are overexposed on the UK
Schmidt Telescope J survey films on which the CSRG is based, but also
due to the camouflage effect of dust within the bar region. Imaging
individual galaxies at NIR wavelengths has a number of advantages in
this regard:

\begin{itemize}
\item The extinction due to dust at 2~\mic\ is only one-tenth that
      at optical wavelengths \citep{ccm89}.
\item The rapid evolution from blue to red supergiants after the first
      5--10~Myr of a starburst \citep{sb99} moves the peak
      of the spectral energy distribution into the infrared. Many of
      these circumnuclear star formation events are not visible at all
      optically in this period.
\item The underlying infrared background is a much better tracer of the
      dominant stellar population in the bar and nuclear region, allowing
      a better estimate of the mass-to-light ratio, and ultimately detailed
      modeling of the gravitational potential which can be compared with
      velocity fields \citep{bc00}.
\end{itemize}

\begin{figure}[h]
\begin{center}
\psfig{file=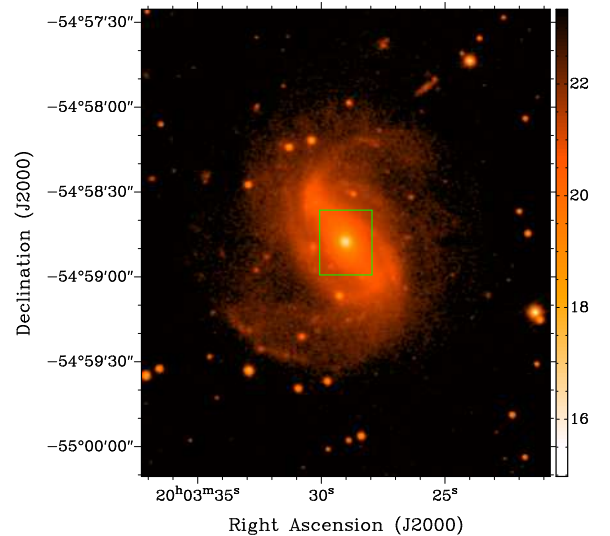,width=3in,bbllx=27pt,bblly=182pt,bburx=537pt,bbury=660pt,clip=}
\caption{IRIS2 $J$-band image of IC~4933. The corresponding surface
brightness in units of mag~arcsec$^{-2}$ is indicated by the colour bar
at right. The inset box outlines the region shown in the colour index maps
in Figures~\protect{\ref{f:j-h}} and \protect{\ref{f:h-k}}.}\label{f:ic4933jsb}
\end{center}
\end{figure}

\begin{figure}[h]
\begin{center}
\psfig{file=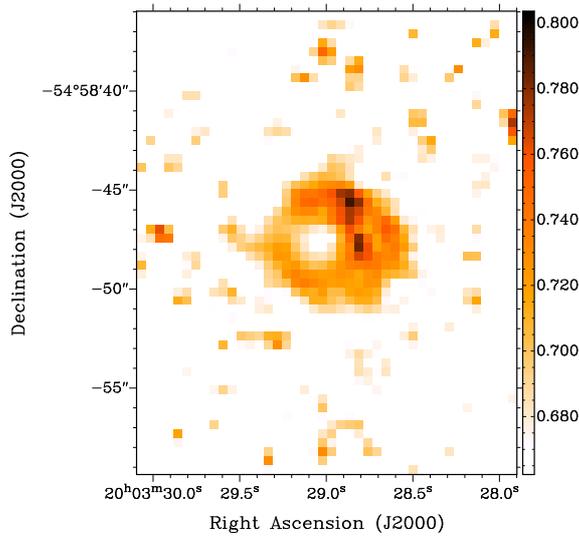,width=3in,bbllx=27pt,bblly=182pt,bburx=537pt,bbury=660pt,clip=}
\caption{IRIS2 ($J-H$) colour index map of the inner bar region of IC~4933.
The colour range displayed is indicated by the scale at right.}\label{f:j-h}
\end{center}
\end{figure}

The existence of a circumnuclear ring in IC~4933 was initially
suspected from the difference of $J$ and $K_{\rm s}$ images obtained
with the Infrared Imager and Spectrograph~2 (IRIS2; \citealt{spie04})
instrument on the 3.9~m Anglo-Australian Telescope (AAT) in June~2003
under marginal conditions. The imaging presented here was obtained on
2006~Sep~4 UT in photometric conditions, with seeing of 1.0~arcsec or
better. The IRIS2 detector is a $1024 \times 1024$ Rockwell HAWAII--1
HgCdTe array with a pixel scale of 0.45~arcsec~pixel$^{-1}$, resulting
in an instantaneous field-of-view (FoV) of $7.7 \times 7.7$~arcmin.

Although the optical extent of IC~4933 is only $\sim$30\% of the IRIS2
FoV, we took a conservative approach and acquired matching
observations of adjacent blank sky to track changes in the background
level and illumination pattern (see \citet{vad04} for a discussion of
the merits of this strategy). Five jittered observations (20~arcsec
offsets) of the target galaxy were bracketed and interleaved with six
jittered observations of the sky 10~arcmin south.  At each object or
sky jitter position, a single 30~sec integration was recorded (in $J$)
or $3 \times 10$~s integrations were averaged (for $H$ or $K_{\rm
s}$). This pattern was repeated twice, for a total on-source exposure
time of 5~min in each of the $J$, $H$, and $K_{\rm s}$ filters.

The data reduction was carried out using the
ORAC-DR\footnote{http://www.oracdr.org/} pipeline within the {\sc
starlink} package and the CHOP\_SKY\_JITTER recipe. Pre-processing of
all raw frames included subtraction of a matching dark frame;
linearity and inter-quadrant
crosstalk\footnote{http://www.eso.org/$\sim$gfinger/hawaii\_1Kx1K/
crosstalk\_rock/crosstalk.html}
correction; and bad pixel masking. The first 6~sky frames are offset
in intensity to a common modal value, then a flatfield is formed from
the median value at each pixel. The 6~sky frames, and 5~object frames
are flatfielded, then the modal pixel values of the two sky frames
bracketing each object frame are averaged and subtracted from that
object frame to account for the sky background. A correction for
astrometric distortion internal to IRIS2 is applied by resampling the
object images, then the actual spatial offsets between images are
computed using point sources common to all images. The 5~object images
are mosaiced together by applying offsets in intensity to the
registered images to produce the most consistent sky value possible in
the overlap regions. A new flatfield and mosaic is constructed for the
second set of 6~sky and 5~object frames, then the two mosaics are in
turn registered and co-added to form a master mosaic. 

Point sources detected in the master mosaic have been cross-referenced
against the 2MASS Point Source Catalog \citep{2mass} and a new
astrometric solution computed, while photometry in a 10~arcsec
aperture has enabled the photometric zero-points to be determined in
each filter, after transforming from the 2MASS $JHK_{\rm s}$ system to
the Mauna Kea Observatories (MKO) $JHK$ system \citep{sdr07}.

\begin{figure}[h]
\begin{center}
\psfig{file=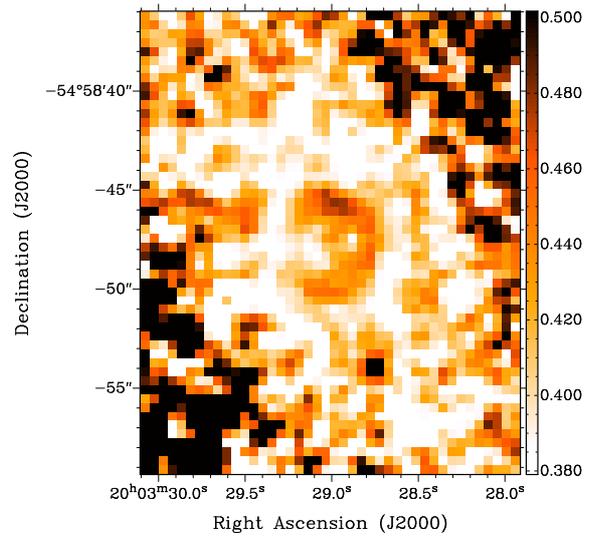,width=3in,bbllx=27pt,bblly=182pt,bburx=537pt,bbury=660pt,clip=}
\caption{IRIS2 ($H-K$) colour index map of the inner bar region of
IC~4933. The colour range displayed is indicated by the scale at right.}
\label{f:h-k}
\end{center}
\end{figure}

Figure~\ref{f:ic4933jsb} shows the $J$-band (1.25~\mic) image of
IC~4933 on a logarithmic scale. The bar, inner ring, and two stubby
spiral arms are evident. The bright compact nucleus is well-fit by a
Gaussian with FWHM of 2.0~arcsec, but shows no obvious indication of
unusual circumnuclear structure. Aligning and subtracting the
calibrated surface brightness images to form colour index maps tells a
different story. Figure~\ref{f:j-h} shows the $(J-H)$ colour of the
central $18 \times 23$~arcsec region, as outlined in
Figure~\ref{f:ic4933jsb}. An almost perfectly circular ring 5~arcsec
(1.5~kpc) in diameter stands out by virtue of its colour; whereas the
underlying stellar population in the bar and bulge has
$(J-H)\sim0.65$, the ring has $(J-H) > 0.7$ and is as red as 0.8 in
the northwest quadrant. The $(H-K)$ map in Figure~\ref{f:h-k}
has less contrast, but confirms the existence of this feature. The
western half of the ring has $(H-K)>0.43$, reaching 0.48 in
the northwest quadrant, while the background colour averages 0.39.

\begin{figure}
\begin{center}
\psfig{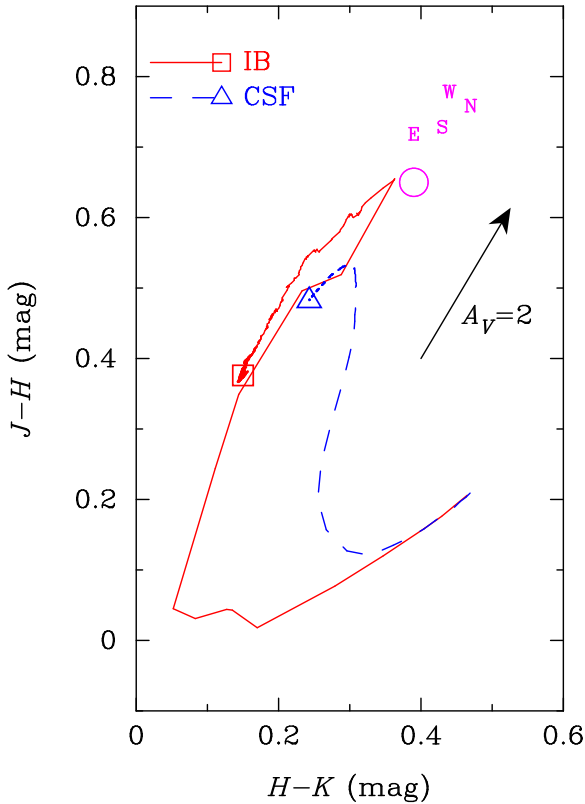}
\caption{Two colour diagram for 4 specific locations around the circumnuclear
ring in IC~4933: (N)orth, (W)est, (S)outh, and (E)ast. The large open circle
indicates the spread of colours from unresolved stars in the bulge and bar.
The arrow indicates the effect on the NIR colours from reddening
equivalent to 2~mag of extinction in the $V$-band. The solid line is the
evolutionary track of a stellar population formed from an instantaneous
burst (IB) according to \citet{sb99}, reaching the open square after
250~Myr. The dashed line ending at the open triangle is the evolutionary
track of a stellar population undergoing continuous star formation (CSF)
over 250~Myr.}\label{f:twocol}
\end{center}
\end{figure}

In Figure~\ref{f:twocol} we plot the $(J-H)$ colour against the
$(H-K)$ colour averaged over a $2 \times 2$~pixel area at the
4~cardinal points around the ring, as well as the mean colours of the
background stellar population. The reddening vector corresponding to
2~mag of extinction in the $V$-band is also plotted. We have used the
Starburst99 models from \citet{sb99} to compute the evolutionary track
through this two colour diagram of a stellar population having an
Initial Mass Function with a Salpeter-like slope ($\alpha=2.35$)
between 1 and 100~\Msun, and super-solar metallicity ($Z=0.04$)
typical of the bulges in Milky Way-like systems such as IC~4933. To
encompass most star formation histories, \citet{sb99} consider two
limiting cases: an `Instantaneous' burst of star formation (IB), and a
`Continuous' constant rate of star formation (CSF). The evolution over
250~Myr for these two scenarios is marked on
Figure~\ref{f:twocol}. Both start off with almost identical colours,
then evolve rapidly bluewards for the first 5--7~Myr until reversed by
the evolution of massive blue stars into red supergiants. The
destruction of these red supergiants in core-collapse supernovae
allows the colours to again evolve bluewards after $\sim$9~Myr and
$\sim$40~Myr for the IB and CSF models, respectively, but now with
$(J-H)$ being 0.5~mag redder than on the initial `blue loop'.

At no stage do these evolutionary tracks achieve colours as red as
those measured around the ring. While the colours of the background
stellar population are similar to those reached by the IB model in the
brief period 8--20~Myr with almost no reddening, this is an extremely
unlikely scenario as the amount of nuclear dust highlighted in
Figure~\ref{f:jhuk} indicates. Rather, being more like the
intermediate age population from continuous star formation marked by
the triangle in Figure~\ref{f:twocol}, the stellar background light in
the bar and nuclear region would experience $A_{V}\sim 2$~mag of
extinction as indicated by the reddening vector. The ring colours are
consistent with an IB population at just 5~Myr old with up to
$A_{V}=6$~mag of extinction, with the required reddening dropping to
0.5--1~mag for an age of 9~Myr. The implied range of extinction from
the CSF models is just $A_{V}=1.5$--2.5~mag, but the allowable range
of ages is much greater at 10--100~Myr. It is this degeneracy between
extinction, age, and even the duration of star formation that requires
us to seek better diagnostics than broadband NIR colours
(Section~\ref{s:gnirs}).

\subsection{Optical Imaging}
\label{s:gmos}

As Figure~\ref{f:twocol} indicates, an alternate scenario is that the
ring is simply dust seen in silhouette, reddening and obscuring the
background population by the equivalent of between 0.3 and 1.4~mag of
extinction in $V$. While linear dust lanes and nuclear dust spirals
are relatively frequent in spiral galaxy nuclei \citep{martini03}, and
compact nuclear dust disks have been seen around some AGN
\citep{n4261,n7052}, such a kiloparsec-scale coherent dust ring has
not been seen elsewhere.

To test for the presence of a dust ring, we need a much larger
extinction contrast than that offered by NIR colours. Images of
IC~4933 in the Sloan Digital Sky Survey (SDSS) $u^{\prime}$,
$g^{\prime}$, and $r^{\prime}$ filters were obtained with the Gemini
Multi-Object Spectrograph (GMOS; \citealt{gmos04}) attached to the
Gemini South Telescope as part of programme GS-2008A-Q-204 (PI:
S. Ryder) in the ``Poor Weather'' queue (seeing $>1.1$~arcsec in
non-photometric conditions with $>2$~mag of atmospheric extinction) on
2008~July~7 UT. Only the $u^{\prime}$ data is presented here. Three
exposures of 690~s each were obtained, with each exposure offset by
8~arcsec spatially to allow filling-in of the inter-CCD gaps in GMOS.
On-chip binning yielded a pixel scale of 0.292~arcsec~pix$^{-1}$.

\begin{figure}[h]
\begin{center}
\psfig{file=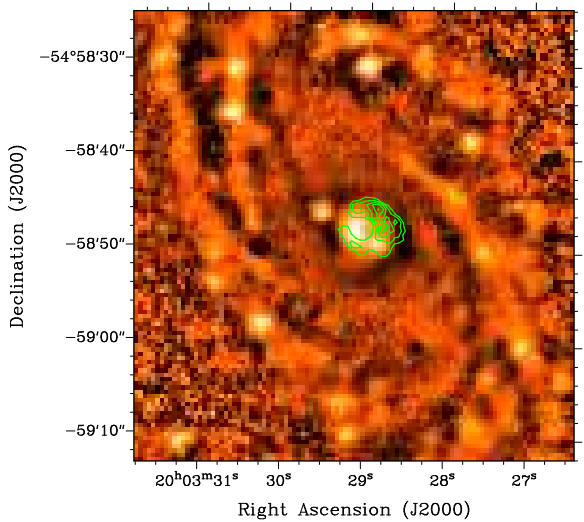,angle=270,width=3.8in}
\caption{Contours of $(J-H)$ colour in the nucleus of IC~4933 from IRIS2,
superimposed on a map of the ($u^{\prime} - K$) colour index in bar
and nuclear region from the difference of GMOS and IRIS2 images. An
unsharp mask has been applied to the ($u^{\prime} - K$) image in order
to highlight the two dust lanes which trace the leading edge of the
bar and encircle the nuclear ring seen in $(J-H)$. The contours
correspond to $(J-H)=(0.71, 0.73, 0.75, 0.77, 0.79)$, while the bluest
($u^{\prime} - K$) colours appear white in this figure.}\label{f:jhuk}
\end{center}
\end{figure}

The data were reduced and combined using V1.9.1 of the {\em gemini}
package within {\sc iraf}\footnote{{\sc iraf} is distributed by the
National Optical Astronomy Observatories, which are operated by the
Association of Universities for Research in Astronomy, Inc., under
cooperative agreement with the National Science Foundation.}. A master
bias frame (constructed by averaging with 3-$\sigma$ clipping a series
of bias frames) was subtracted from all raw images in lieu of overscan
fitting and subtraction. Images of the twilight sky were used to
flatfield the images, then the dithered galaxy images were registered
and averaged together with the {\em imcoadd} task to eliminate the
inter-CCD gaps, bad pixels, and cosmic rays. No absolute flux
calibration of this data is available, but for the purposes of
constructing colour index maps the final image has been converted
to a logarithmic scale.

This $u^{\prime}$ image has been scaled and aligned to match the IRIS2
$K$ image using the {\em geomap} and {\em geotran} tasks before the
latter was subtracted from the former to produce a ($u^{\prime} - K$)
colour index map which spans a factor of 6 in wavelength. This map has
been unsharp masked by subtracting off a lightly smoothed version of
itself, to help accentuate any apparent dust lane
structure. Figure~\ref{f:jhuk} shows this difference in optical versus
NIR morphology, together with a contoured representation of the ring
seen in $(J-H)$. One can quite clearly see dust lanes offset either
side of the bar major axis and slightly concave towards it. As shown
in the simulations of \citet{ath92}, such offset curved dust lanes
mark the loci of shocks at the leading edge of a relatively fat bar.
These dust lanes curl around the nucleus and encircle it, though this
dust ring is noticeably larger ($\sim8$~arcsec diameter) than the NIR
ring found in Section~\ref{s:iris2}. Within this dust ring sits three
blue `hot spots', two of which are concentric with the NIR ring and
which in fact coincide with the bluest $(J-H)$ sections of the NIR
ring.

\subsection{IFU Spectroscopy}
\label{s:gnirs}

Near-IR spectroscopy can place much tighter constraints on the ages
and star-forming timescales of individual clusters
\citep{rkt01,ark01}. The combination of multiple, independent age diagnostics
(including the \brg~2.16~\mic\ equivalent width; the
\co~2.3~\mic\ spectral index; or the \fe~1.26~\mic\ luminosity) can in
principle discriminate the IB from the CSF models
of \citet{sb99}, as well as their relative ages. Since most of these
diagnostics involve the ratio of an emission/absorption feature to
their adjacent continuum emission, they are far less affected by
differential extinction than the broadband colour indices in
Section~\ref{s:iris2}. 

Spectroscopy of the circumnuclear ring in IC~4933 was conducted in
queue mode with the Gemini Near-InfraRed Spectrograph (GNIRS;
\citealt{gnirs}) with its Integral Field Unit (IFU; \citealt{ifu})
attached to the Gemini-South Telescope for program GS-2005A-Q-25 (PI:
S. Ryder). The GNIRS IFU slices a $4.8 \times 3.15$~arcsec FoV into
21~slices 0.15~arcsec wide by 4.8~arcsec long and reformats it into a
single `long slit', with each of the 21~slices projecting to 32
0.15~arcsec pixels on the $1024 \times 1024$ Aladdin InSb array. The
31.7~lines~mm$^{-1}$ grating and short camera disperse the stacked
slices at a resolving power $\lambda / \Delta \lambda \sim 1700$, with
separate grating settings spanning much of the $J$-band
(1.06--1.45~\mic) and the $K$-band (1.90--2.56~\mic).

As the ring just overfills the IFU FoV, four adjacent pointings were
observed, effectively covering each quadrant of the ring in turn. Each
pointing was observed in $4 \times 250$~sec integrations, with
32~reads per integration delivering a read noise of 7~electrons. Each
of the 4~integrations was dithered by $\pm$0.3~arcsec (2~pixels) in
both axes to compensate for reduced throughput in the two outer and
one of the central slices, and to provide 2~pixels of overlap between
pointings. Furthermore, each integration was interleaved with
equivalent integrations of blank sky 1~arcmin east, to enable removal
of the telluric OH and thermal emission. Observations of the F0~V
stars HD~176193 and HD~200767, nodding between each half of the IFU,
were made at comparable airmass to the IC~4933 observations to correct
for telluric absorption. Observations were conducted over 5~nights in
March and April 2005 when conditions were photometric, in seeing of
0.8--1.0~arcsec. Spectra of an Argon arc lamp provided the wavelength
calibration, while observations of a 1100~K `grey body' or quartz
halogen lamp facilitated the correction for pixel-to-pixel sensitivity
variation.

The IFU data has been reduced using the {\em gnirs} package within
{\sc iraf}. Images taken in `Bright Objects' read mode (i.e., telluric
standards, flatfields, and arcs) were found to exhibit a ripple
pattern noise along each column, having a period of 17~pixels and
amplitude $\sim$260~electrons, well in excess of the 38~electrons read
noise expected in this mode. We have been able to remove this ripple
pattern by median filtering the unexposed columns, then subtracting
this from every column in the image. The 32~columns corresponding to
each of the 21~slices are extracted into individual image planes. Each
of the 21~object slices has the matching nodded object or blank sky
slice subtracted from it, and is then divided by the normalised,
smoothed flatfield lamp slice. Tracing of an artificial `pinhole'
spectrum allows each slice plane to be straightened in the spatial
direction, while interactive arc line identification and fitting is
used to rebin each slice on a linear wavelength scale.

Subsequent processing has been carried out using purpose-written {\sc
idl} routines. For the IC~4933 observations the 21~slices are stacked into
a 3D datacube of $32 \times 21$ spatial pixels $\times$~1022 spectral
elements. The 4~datacubes from each of the dithered pointings covering
each quadrant are registered and combined into one datacube per
quadrant. The 4~quadrant cubes have then been mosaiced into a master
datacube covering the entire ring and its surroundings.

For the stellar observations, columns containing high signal-to-noise
spectra are extracted and summed; Gaussian profiles are fitted to the
hydrogen recombination lines (e.g., Pa$\beta$ at 1.28~\mic, \brg\ at
2.16~\mic) then used to fill in this absorption; and finally they are
divided by the Planck function appropriate to a blackbody of the same
temperature as the star. This yields response spectra which are
divided into each spectrum of the $J$-band and $K$-band master
datacubes to compensate for variations in atmospheric transmission
with wavelength.

\begin{figure}[h]
\begin{center}
\psfig{file=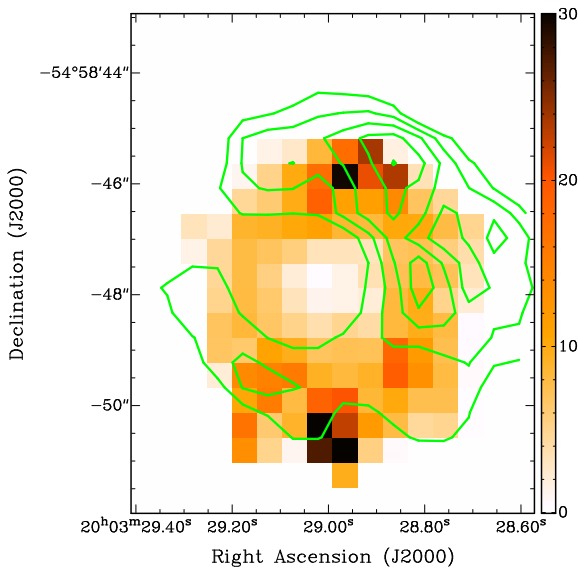,angle=270,width=3in,bbllx=27pt,bblly=150pt,bburx=568pt,bbury=670pt,clip=}
\caption{Contours of $(J-H)$ colour in the nucleus of IC~4933 from IRIS2,
superimposed on a map of the \pab\ equivalent width from GNIRS IFU
spectroscopy. The contours correspond to $(J-H)=(0.71, 0.73, 0.75, 0.77,
0.79)$, while the equivalent width in units of \AA\ is indicated by the scale
at right.}\label{f:pbjh}
\end{center}
\end{figure}

Inspection of the datacubes does indicate the presence of extended
line emission at the redshifted wavelengths of \pab, \brg, and \h2.
However, the net signal-to-noise in the $K$-band master datacube is
significantly poorer than the $J$-band, and inadequate for our
purposes. We therefore have to rely solely on the $J$-band datacube
for our analysis. The equivalent width of the \pab\ emission line has
been measured in the datacube by integrating the flux within the line
and dividing by the interpolated continuum level. Figure~\ref{f:pbjh}
shows a map of the \pab\ equivalent width after binning $\times3$ to
match the pixel scale of IRIS2, overlaid with contours of the $(J-H)$
colour from Figure~\ref{f:j-h}.

\begin{figure}[h]
\begin{center}
\psfig{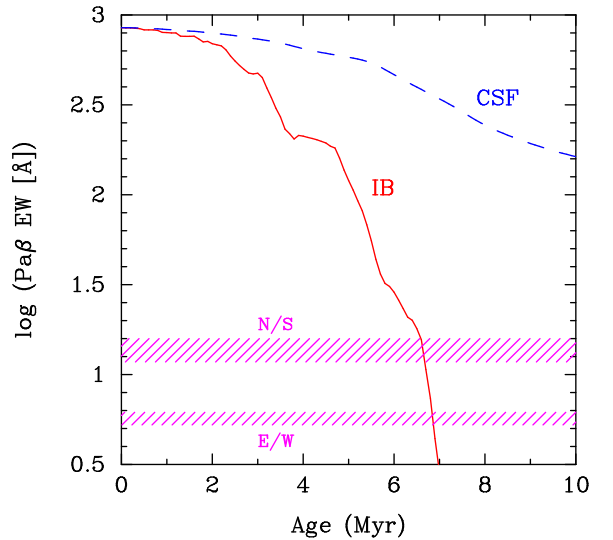}
\caption{The evolution with time of the \pab\ equivalent width for
the same Starburst99 IB and CSF models as in Fig.~\protect{\ref{f:twocol}}.
The upper and lower shaded regions mark the observed range of equivalent
widths within the North/South, and East/West quadrants respectively, of the
IC~4933 nuclear ring.}\label{f:pbage}
\end{center}
\end{figure}

By contrast with the broad-band colour maps, the \pab\ map shows a
clear bifold symmetry, with the equivalent width reaching a maximum of
$14\pm2$~\AA\ almost due North and South around the ring, and a
minimum of $5.5\pm0.5$~\AA\ in the East and West. Figure~\ref{f:pbage}
shows the evolution of \pab\ equivalent width with time for the IB and
CSF Starburst99 models of Section~\ref{s:iris2}.

The IB model reaches the \pab\ levels of the North/South quadrants
6.6~Myr after the onset of the burst, and those of the East/West
quadrants a further 0.3~Myr later. This is perfectly consistent with
the IB colour evolution described by Fig.~\ref{f:twocol}, with
azimuthal variations in extinction accounting for the morphological
differences between the nuclear ring as seen in $(J-H)$ and in \pab\
(Fig.~\ref{f:pbjh}).  Indeed, the extinctions inferred for each
quadrant from their derived ages and location in Fig.~\ref{f:twocol}
are found to range from $A_{V}
\sim 2$~mag in the E~quadrant, up to 3.5~mag in the N~quadrant.
%

Although this plot covers only the first 10~Myr, even 1~Gyr after the
onset of CSF the \pab\ equivalent width is still 50\% higher than the
peak values observed anywhere around the ring. While it is conceivable
that the observed \pab\ equivalent widths are underestimated due to
inadequate allowance for an older background stellar population
contaminating the continuum fluxes, the very gradual decline in \pab\
equivalent width over time in the CSF model would make the East/West
quadrants at least 0.5~Gyr older than the North/South quadrants. We
can conceive of no scenario under which two populations of such
differing ages could remain spatially distinct for so long in such a
compact, but kinematically active region and therefore discount purely
continuous star formation in the circumnuclear ring of IC~4933 as
playing any significant role.

\section{Discussion}
\label{s:disc}

The IB and CSF Starburst99 models represent the two extremes of star
formation, and as stellar population model fits to optical line
indices of nuclear rings by \citet{sakm07} has indicated, the truth is
more likely to be a series of multiple, episodic bursts superimposed
on an old underlying bulge population. Nevertheless, the age
differences around the ring are unmistakable. IC~4933 thus joins a
small, but growing list of galaxies whose nuclear rings exhibit
evidence of a clear bipolar azimuthal age gradient such as M100
\citep{rkt01,all06}, NGC~1343, and NGC~1530 \citep{lmm08}. Many more
galaxies show some form of azimuthal age gradient at least, like that
in M83
\citep{pdw97,har01,jhk10}, including 60\% of the sample of
\citet{bok08} observed in the NIR with the SINFONI IFU on the VLT,
and 40\% of the sample of \citet{lmm08} observed in H$\alpha$.

\citet{lmm08} found no clear relationship between the presence or
absence of an age gradient, and the morphology of either the ring or
the host galaxy. They did note that the rings in their sample that
show no clear age gradient are also among the oldest ($>$10~Myr)
star-forming rings. Thus it may be that with the passage of time and
several complete orbits of the aging clusters around the ring that a
well-organised, sequential star formation pattern may simply be
smoothed over \citep{bok08}. We do know that the starburst in IC~4933
is comparatively young and of short duration, which may partly account
for why it still has such a well-defined bipolar age gradient. A weak
or intermittent gas inflow rate from the bar into the ring may also
yield a more stochastic age distribution of clusters around the ring,
particularly if star formation only occurs above some critical
threshold density \citep{bge97,lmm08,bok08}.

\citet{lmm08} plotted the angular offset between the bar major axis
position angle and the azimuth of the youngest H\,{\sc ii}~region
around each half of their nuclear rings. They found a broad
distribution with an apparent peak around $90^{\circ}$, suggesting
that soon after inflowing gas reaches the ``contact points'' between
the offset dust lanes and the nuclear ring (commonly found
perpendicular to the bar major axis) it is converted to stars.  We
have used the {\em ellipse} surface photometry task within {\sc iraf}
to derive the bar position angle. Over radial distances between 10 and
18~arcsec where the bar dominates, the position angle is
$(32\pm2)^{\circ}$. The two \pab\ peaks in Figure~\ref{f:pbjh} have
position angles of $182^{\circ}$ and $350^{\circ}$, so are displaced
some $30^{\circ} - 40^{\circ}$ from the bar position angle. The
spatial resolution of Figure~\ref{f:jhuk} is insufficient to discern
if the contact points in IC~4933 are indeed at $90^{\circ}$ to the bar
major axis, or could in fact be much closer to these youngest H\,{\sc
ii}~regions.

The GNIRS \pab\ datacube allows us to extract a crude velocity field
which yields a deprojected rotation velocity around the ring of
220~\kms\ with the northeast side of the disk approaching us.
Assuming a bar pattern speed in the region of 25~\kms~kpc$^{-1}$
\citep{ryd96}, newly-formed clusters would then drift away from the
contact point at $\sim$200~\kms. At that velocity, one complete orbit
of the ring takes just 23~Myr. So we should perhaps not be too
surprised to find relatively young clusters so far from the inferred
current location of the contact points, where star formation could be
happening right now but will not become apparent even in the NIR for a
few Myr.

Alternatively, as shown by the simulations of \citet{hs96}, an oval
(rather than circular) nuclear ring can result in contact points which
are not perpendicular to the bar major axis. It is notable that the
nuclear ring seen in the colour maps and in \pab\ is quite circular,
whereas IC~4933 appears to be inclined to the line-of-sight by
$37^{\circ}$ \citep{lv89}. Thus, foreshortening of an eccentric
nuclear ring could result in it appearing circular in the plane of the
sky, and it may in fact be that the contact points of the dust lanes
with the ring do coincide with the youngest H\,{\sc ii}~regions.


Interestingly, the blue `hot spots' in the ($u^{\prime} - K$) image
(Figure~\ref{f:jhuk}) also show large spatial offsets from the
\pab\ equivalent width peaks. The Starburst99 IB model in
Figures~\ref{f:twocol} and \ref{f:pbage} predicts (in the absence of
any reddening) that ($U-K$) becomes bluer by 1~mag in the first
$4-5$~Myr, then reddens rapidly by 4~mag in the next 3~Myr and stays
within 1~mag of this thereafter. For these hot spots to appear as blue
as they do, in the presence of dust, would require that they be no
older than 5~Myr. This is inconsistent with the chronology derived
from \pab\ equivalent widths, which are far less susceptible to
reddening, and again highlights the risk in relying solely on
broadband colour measurements for deriving star formation histories on
short ($<$10~Myr) timescales.

\section{Conclusions}

Motivated by the fact that barred spiral galaxies known to possess
large-scale inner rings are more likely to host nuclear rings as well,
we obtained $J$, $H$, and $K_{\rm s}$ imaging of the nearby SB(rs)bc
galaxy IC~4933 with IRIS2 on the AAT, and $u^{\prime}$ imaging with
GMOS on Gemini South. Although the reduced images show only the bulge,
bar, and arms, an almost circular ring 5~arcsec (1.5~kpc) in diameter
is seen in the colour index maps. Comparison of the ring NIR
colours with evolutionary tracks from Starburst99 models is unable to
distinguish between a burst population just 5~Myr old, and
less-reddened continuous star formation over the past 10--100~Myr.

The equivalent width of the \pab\ line is much less susceptible to
reddening effects. We have mosaiced four pointings of the GNIRS
Integral Field Unit in order to map the \pab\ equivalent width around
the ring.
We find clear peaks in the equivalent width in the northern and
southern sectors which are most easily accounted for by short-lived
bursts of star formation 6.6~Myr ago rather than by continuous star
formation, with the eastern and western sectors appearing to be
0.3~Myr older. IC~4933 is the fourth galaxy seen to have evidence of
such a well-defined bipolar azimuthal age gradient. Although the
actual location of the contact points of the offset dust lanes in the
bar with the nuclear ring cannot be clearly discerned in our images,
the significant azimuthal offset between the youngest H\,{\sc
ii}~regions in the ring and the bar major/minor axes are consistent
with this nuclear ring being intrinsically oval in shape but
fortuitously appearing circular from our perspective. Ongoing and
future NIR imaging surveys such as those conducted with UKIRT and
VISTA can be expected to yield many more such examples of nuclear
rings, but only a comprehensive NIR integral field spectroscopic
followup can unambiguously define the star formation history around
these rings.

\section*{Acknowledgments}

Based in part on observations obtained at the Gemini Observatory,
which is operated by the Association of Universities for Research in
Astronomy, Inc., under a cooperative agreement with the NSF on behalf
of the Gemini partnership: the National Science Foundation (United
States), the Science and Technology Facilities Council (United
Kingdom), the National Research Council (Canada), CONICYT (Chile), the
Australian Research Council (Australia), Minist\'{e}rio da Ci\^{e}ncia
e Tecnologia (Brazil) and Ministerio de Ciencia, Tecnolog\'{i}a e
Innovaci\'{o}n Productiva (Argentina). This publication makes use of
data products from the Two Micron All Sky Survey (a joint project of
the University of Massachusetts and the Infrared Processing and
Analysis Center/California Institute of Technology), as well as the
NASA/IPAC Extragalactic Database (NED, which is operated by the Jet
Propulsion Laboratory, California Institute of Technology), funded by
the National Aeronautics and Space Administration and the National
Science Foundation. We thank the referee for some helpful suggestions.


\end{document}